# Formation of martensitic microstructure in epitaxial Ni-Mn-Ga films after fast cooling

Yuru Ge[1,2,*], Fabian Ganss[1], Klara Lünser[1], Satyakam Kar[1,2,3], René Hübner[1], Shengqiang Zhou[1], Lars Rebohle[1], and Sebastian Fähler[1]

[1]Helmholtz-Zentrum Dresden-Rossendorf, Institute of Ion Beam Physics and Materials Research, 01328 Dresden, Germany

[2]TU Dresden, Faculty of Mechanical Science and Engineering, 01062 Dresden, Germany

[3]Leibniz IFW Dresden, Institute for Metallic Materials, 01069 Dresden, Germany

*Corresponding author: Yuru Ge, y.ge@hzdr.de

**Abstract**

Shape memory alloys have a wide range of applications, including high stroke actuation, energy-efficient ferroic cooling, and energy harvesting. These applications are based on a reversible martensitic transformation, which results in a complex microstructure in the martensitic state. Understanding the formation of this microstructure after fast heating and cooling is crucial, as a high cycle frequency is essential for achieving high power density in the mentioned applications. This article uses an epitaxial Ni-Mn-Ga film as a model system to study the formation of the martensitic microstructure, since the high surface-to-volume ratio of thin films enables rapid heating and cooling, and since the formation of a multi-level hierarchical microstructure during slow cooling of this material system is already well understood. We apply a millisecond flash lamp pulse of varying energy density on the Ni-Mn-Ga film and analyse the hierarchical martensitic microstructure afterwards. Substantial differences compared to slow cooling occur, which we attribute to the limited time available for the microstructure to form and to the thermal stress between film and substrate during rapid temperature changes.

Keywords: Ni-Mn-Ga thin film; epitaxial growth; martensitic transformation; microstructure; flash lamp annealing.

1. Introduction

The martensitic transformation (MT) of shape memory alloys makes these functional materials susceptible to stress and strain, which enables high stroke, high force actuation [1]. Furthermore, MTs facilitate emerging applications like energy-efficient ferroic cooling [2] and waste energy harvesting [3]. During an MT cycle, the material switches reversibly between the high-temperature austenite (A) and the low-temperature martensite (M) phases. An MT is diffusionless, which makes it faster than conventional structural phase transformations that require atomic diffusion. Applications benefit from a high cycle frequency, as their power density is proportional to frequency. However, the majority of the research on MT covers only conventional slow heating and cooling cycles, which take several seconds up to minutes. The influence of fast transformation has only been examined sparsely [4-11]. The most systematic studies on fast transformations are reported by the group of Shilo [4-6, 10, 11]. For example, they examined 0.2 mm and 0.38 mm diameter NiTi wires in a unique setup with Joule heating to induce a reverse MT from martensite to austenite with a 5 µs electric pulse and probed the mechanical response down to the microsecond time scale [10]. This setup was further enhanced with time-resolved X-ray diffraction, revealing that MT occurs near the surface of the wire after 9 µs, which is faster than within the wire, where MT starts after approximately 40 µs [11]. However, all examples above refer to bulk wires, which exhibit low surface-to-volume ratios and thus are unfavourable for faster cycles, as the heat exchange must proceed through the surface.

Thin films with their large surface-to-volume ratio can achieve faster MT and are, therefore, well suited for a high cycle frequency. So far, only few experiments on fast transformation in thin films were published [12-14]. For instance, films were heated with a short laser pulse to induce a fast transformation [9, 12]. These experiments revealed that a reverse transformation can be achieved within 7 ns, and a cycle could be completed within several 100 ns as heat is quickly transferred into the substrate [14]. However, these experiments only probed the dynamic crystal structure at the atomic scale. Up to now, studies on the martensitic microstructure formation after a fast MT are missing. This is a substantial shortcoming, as the martensitic microstructure, which covers a much broader length scale and consists of many different types of twin boundaries, decides on most functional properties of shape memory alloys.

To study the formation of martensitic microstructure obtained by fast transformation, flash lamp annealing (FLA) is selected, since it allows fast heating within hundreds of microseconds to tens of milliseconds [15]. We use epitaxially grown Ni-Mn-Ga films as a model system, as they exhibit a well-ordered hierarchical martensitic microstructure in the as-deposited state [16-18]. For the present experiments, we employ 500-nm-thick Ni-Mn-Ga films on single-crystalline MgO (001) substrates and take the as-deposited martensitic microstructure as a reference state. To reach a fast cooling rate through the MT, we chose films with a high martensitic start temperature ($M_s$) of about 610 K. We heat the as-deposited films with a single FLA pulse at different energy densities ($E_D$). To investigate the crystal structure and martensitic microstructure at all length scales, we combine reciprocal space mapping (RSM), focused ion beam (FIB) cross-sectioning with scanning electron microscopy (SEM), and (scanning) transmission electron microscopy ((S)TEM) with spectrum imaging analysis based on energy-dispersive x-ray spectroscopy (EDX). We show that these complementary methods allow to quantify changes of the martensitic microstructure by FLA for the different types of twinning. To understand the influence of $E_D$, we present simulations of the temperature evolution during FLA and derive three mechanisms, through which the FLA pulse influences the microstructure formation. In our discussion, these three mechanisms allow to explain the observed changes at each level of the hierarchical microstructure. However, for each level, the underlying processes happening inside the material are different. As our work covers most of the levels of twinning, we obtain an idea of the dynamics relevant for all processes forming a martensitic microstructure from the atomic scale up to the macroscale.

## 2. Reference state: hierarchical microstructure with five levels of twinning

To study the formation of the martensitic microstructure achieved after a fast transformation by FLA, it is necessary to first describe the martensitic microstructure of the as-deposited state. After depositing the films at high temperature in the austenitic state, these reference samples are cooled to the martensitic state at room temperature within minutes (details are described in Subsection 7.1). Following this relatively slow cooling, a hierarchical microstructure with five levels of twinning, ranging from the nano- to the macroscopic scale, is observed [16]. The construction of this hierarchical twinning is sketched in *Figure 1*a.

The starting point is the tetragonal martensitic unit cell, which forms from the cubic austenite via a MT. This tetragonal structure is assigned as non-modulated martensite (NM). Nano-twin boundaries (nano-TBs, orange lines at level 1) are introduced between differently oriented tetragonal building blocks, as this is the major process to minimize the elastic energy at the habit plane which connects austenite and martensite. Accordingly, this twinning occurs at level 1, at the nanoscale. In addition to NM, modulated martensites (MM, e.g. 10M and 14M) occur, too. Following the adaptive concept [19, 20], these phases are constructed from NM by inserting nano-TBs in a specific periodicity. In the present 500 nm-thick film, we observe 14M [18]. There are two orientations to form 14M periodic modulations by introducing nano-TBs: $(5\bar{2})_2$-stacking or $(2\bar{5})_2$-stacking. Connecting these two variants by mirroring results in a so-called (a/b)$_{MM}$-TB at level 2 (green line in *Figure 1*a). The process of introducing (a/b)$_{MM}$-TBs is similar to level 1. It allows to minimize the elastic energy at the habit plane further, beyond what is achievable with a single 14M variant [18]. Further away from the habit plane, where energy minimization through a close spacing of TBs is not critical, the coarsening of TBs can lead to a transition between MM and NM [21].

The next level of twinning is required for nucleation and growth of martensite, an essential process for a first-order MT. At level 3, (a/b)$_{MM}$-laminates of different orientations are used as building blocks for diamond- and parallelogram-shaped nuclei of martensite [17]. These shapes are possible because the (a/b)$_{MM}$-laminates are compatible with austenite at the outside, and inside the nuclei. Additional types of mesoscopic TBs form between the (a/b)$_{MM}$-laminates, in particular type-Ⅰ TBs and type-Ⅱ TBs (as marked in level 3 in Figure 1a). Typically, the first nucleus creates many parallel nuclei by so-called autonucleation, resulting in laminates of mesoscopic twin boundaries, also known as colonies [22]. Since type-Ⅱ TBs deviate slightly from a precise (101)$_A$ plane, two different orientations of the nuclei can form, which are connected by a so-called herringbone TB at level 4 (this process is not examined here). Additionally, one has to consider that six different orientations of {110}$_A$ planes are possible [23, 24]. Accordingly, six fundamentally different orientations of colonies form, and this process results in boundaries between these colonies, referred to as macroscopic TBs at level 5. In the present case of thin films, not all 6 orientations are equivalent anymore, and thus, one can distinguish two types easily by their traces on the surface [17, 23]: type-X consists of TBs inclined by 45° within the substrate plane (with high contrast and traces parallel to ⟨100⟩$_A$ on the film surface), and type-Y with TBs parallel to the substrate edges (with low contrast and traces parallel to ⟨110⟩$_A$).

FLA affects most of the five microstructure levels, as summarized in Figure 1b. This summary is based on the experimental observations described in Section 3 and interpreted in Section 5. To confirm exemplarily that the present films exhibit the same hierarchy of twinning and to illustrate level 3…5 experimentally, we show an SEM image in Figure 2a. In this top view, one can clearly observe level 5 with the two types of colonies: type-X (dark area) and type-Y (bright area). A zoom into the area marked by the black rectangle (Figure 2b) shows that type-X and -Y colonies are connected by macroscopic TBs. They can be distinguished clearly due to their difference in contrast. As an example, one type-X colony is marked by a dark shadow and one macroscopic TB in brown. A further enlarged image in Figure 2b reveals some parallelogram laminates at level 3.

### 3. Experimental observations

To investigate the changes in crystal structure and hierarchical microstructure after FLA at different energy densities ($E_D$), this section compares the observations with the reference state, starting with the atomic scale up to the macroscale. Details on the methods used are given in Section 7. To keep this paper concise, we show the detailed results of the as-deposited film and the three FLA-treated films with $E_D$ = 28, 55, and 83 J/cm$^2$ only. Results on all other $E_D$ (14, 42, 69 and 76 J/cm$^2$) are presented in the supplementary. Interpretations are given in corresponding subsections in Section 5.

**3.1 At the nanoscale: lattice parameters and phase intensity fractions**

To characterize the crystal structure at level 0…2 at the nanoscale, reciprocal space maps (RSM) were recorded. (Details about RSM measurements are given in Supplementary Figure S1.) The RSMs of the as-deposited film and the three FLA-treated films with $E_D$ = 28, 55, and 83 J/cm$^2$ are shown in Figure 3. The other four RSMs are given in Supplementary Figure S2. Reflections are indexed as tetragonal NM (($400$)$_{NM}$ and ($004$)$_{NM}$ with blue dashed boxes) and orthorhombic 14M (($400$)$_{14M}$, ($004$)$_{14M}$, and ($040$)$_{14M}$ with green dashed boxes). The reflections match well with the positions expected from the literature [20]. In addition, a substantial ($004$)$_A$ reflection (marked with a red dashed box) indicates the presence of some residual austenite in the as-deposited state. After FLA, the following changes are observed with increasing $E_D$: the intensities of the ($040$)$_{14M}$ and ($004$)$_{14M}$ reflections decrease, while changes of ($400$)$_{14M}$ are imperceptible. Correspondingly, the intensities of ($400$)$_{NM}$ and ($004$)$_{NM}$ increase. Regarding the residual austenite, the intensity of the (004)$_A$ reflection decreases with increasing $E_D$.

To quantify these trends, the lattice parameters, intensity fractions, and coherence lengths (*D*) were extracted from the RSMs and are summarized as a function of $E_D$ in Figure 4. The lattice parameters of cubic A, tetragonal NM, and orthorhombic 14M, as shown in Figure 4a, remain unchanged after FLA. In some samples, the intensities (Figure 4b) vanished, and accordingly, no values for the lattice parameters could be determined. Indeed, the intensities of the three phases vary strongly as a function of $E_D$ and are plotted in Figure 4b after normalization to the total intensity. The as-deposited film consists of a high fraction of 14M, NM, and some A residue. Up to $E_D \leqslant 42$ J/cm$^2$, as illustrated by the light grey background, the intensity fractions of NM and 14M change only slightly. With $E_D > 42$ J/cm$^2$, as marked by the light orange background, the intensity fraction of the 14M reflections decreases and that of the NM reflections increases accordingly. The intensity of residual austenite decreases and disappears for $E_D > 55$ J/cm$^2$. However, due to the overlap of (004)$_A$ and (040)$_{14M}$, a minor fraction might still be present but cannot be quantified. In addition, the coherence lengths of four sets of lattice planes were calculated by the Scherrer equation (Details in Supplementary Figure S1) and are summarized in Figure 4c. After FLA with $E_D \leqslant 42$ J/cm$^2$, the coherence lengths of both NM and 14M decrease slightly. However, if $E_D$ is larger than 42 J/cm$^2$, an increase of $D_{(400)14M}$ and $D_{(040)14M}$ and a strong increase of $D_{(004)NM}$ is observed. To sum up, we observe that if $E_D$ exceeds 42 J/cm$^2$, the NM fraction increases while the 14M phase decreases correspondingly, accompanied by an increase in the coherence lengths. The interpretation of these trends is provided in Subsection 5.1.

**3.2 At the mesoscale: type-X twinning period ($\Lambda_{TB}$)**

To examine the influence of FLA on the microstructure at level 3 at the mesoscale, Figure 5 and Supplementary Figure S3 present representative SEM micrographs of the as-deposited film and after FLA-treatment with various energy densities $E_D$. We focus on the type-X microstructure as its sawtooth topography generates much better contrast than type-Y variants [25, 26]. This high contrast allows to resolve the nuclei. Due to the epitaxial relationship, in these images, traces of type-X laminates are aligned at 45° with respect to the image edges (<100>$_{MgO}$ || <110>$_{Ni-Mn-Ga}$). Compared to the as-deposited film (Figure 5a), no noticeable difference is visible up to $E_D \leqslant 42$

J/cm² (Figure 5b and Figure S3a-b). However, at higher $E_D$, the period of the mesoscopic TBs ($\Lambda_{TB}$) appears to be reduced. To quantify this observation, $\Lambda_{TB}$ was determined for each $E_D$ from five SEM images (as those given in Figure 5 and Supplementary Figure 3), which gives sufficient statistics. As summarized in Figure 6a, $\Lambda_{TB}$ remains almost constant around 115 nm for $E_D < 42$ J/cm². At higher $E_D$, $\Lambda_{TB}$ decreases linearly down to 85 nm at $E_D = 83$ J/cm². The interpretation of the linear decrease of $\Lambda_{TB}$ for $E_D$ is above 42 J/cm² is found in Subsection 5.2.

**3.3 At the macroscale: occurrence of small colonies and transition from type-Y to -X**

A detailed examination of the micrographs in Figure 5 and Supplementary Figure S3 reveals that many small colonies appear after FLA treatment at $E_D \geqslant 42$ J/cm² (marked by white dotted contours). Following the concepts described in Section 2, colonies are connected by macroscopic twin boundaries at level 5. To quantify the influence of FLA on colonies, five SEM micrographs with a total image area of 1322 µm² were analysed (see accompanying raw data). Figure 6b summarizes the size distribution of colonies, which fall into four categories: <1, 1…10, 10…100, and > 100 µm². While the size distribution of the as-deposited film and the one annealed at 28 J/cm² is quite similar, a significant increase in the number of smaller colonies is observed for $E_D$ of 55 and 83 J/cm².

To probe the influence of FLA on type-X and type-Y variants at level 5, SEM analyses were performed at a much lower magnification than for the colony study. Representative images are shown in Figure 7 and in Supplementary Figure S4. At this magnification, type-X and type-Y variants are easily distinguishable due to their different contrasts. In the as-deposited film, the martensitic microstructure shows approximately equal surface fractions of type-X and type-Y patterns (Figure 7a). With increasing $E_D$, the microstructure changes significantly with the area of type-X increasing at the expense of type-Y (Figures 7b-c and Figures S4a-c). For $E_D = 76$ and 83 J/cm², type-Y colonies almost completely disappear (Figure S4d and Figure 7d). To quantify this observation, five SEM images were analysed for each $E_D$. As summarized in Figure 8, for the as-deposited film, the fractions of type-X and type-Y ratios are almost equal to 50 %. After FLA, a linear decrease of type-Y to the advantage of type-X is observed. At $E_D = 83$ J/cm², type-Y almost disappears. This quantitative analysis confirms the initial impression from Figure 7 and Figure S4. As an additional independent method, XRD pole figure measurements were used (see accompanying raw data) to distinguish types-X and Y due to their different tilt and rotation (see supplementary Figure S5). In contrast to SEM, which probes essentially the sample surface, XRD probes the whole volume of these 500-nm-thick films. Both independent methods give similar results and show the same trend as a function of $E_D$. This agreement also indicates that type-X and -Y twin boundaries extend through most of the film, from the surface to the substrate. To sum up this subsection, a significant increase in the number of smaller colonies is observed for $E_D$ of 55 and 83 J/cm$^2$. In addition, the variant fraction of type-X increases linearly with the increasing of $E_D$. These observations are interpreted in Subsection 5.3.

### 3.4 At the macroscale after very high $E_D$: wrinkles and sample cracking

After FLA at $E_D$ = 76 and 83 J/cm$^2$, many dark and bright traces parallel to both substrate edges (|| ⟨100⟩$_{MgO}$) are observed (Supplementary Figure S4d and Figure 7d, respectively). To better understand the origin of these traces, an enlarged image of the film with $E_D$ = 83 J/cm$^2$ is shown in Figure 9a, According to this SEM micrograph, they are film wrinkles with two features: valleys and hillocks. A FIB-prepared cross-section through one narrow hillock is given in Figure 9b. In this hillock region (marked by a black rectangle), the film still appears to be well attached to the substrate with a bump at the interface to the substrate. For a more detailed analysis at even higher resolution, a cross-sectional TEM lamella was prepared through a narrow hillock region (see Supplementary Figure S6a, b). In addition to a slight increase in film thickness, the bump at the interface to the MgO substrate is clearly visible. However, no delamination of the Ni-Mn-Ga film is observed.

To investigate whether FLA at such high energy density affects the element composition, high-angle annular dark-field scanning transmission electron microscopy (HAADF-STEM) imaging and spectrum imaging analysis based on EDX were performed for the specimen region shown in Figure S6a (See Supplementary Figure S6c-j). In particular, the area marked by the green box in Figure S6c was analysed. While some smaller film regions, which are not restricted to the bump area, deviate slightly from the average composition (Figure S6d), the overall element distributions are well maintained (Figure S6e - j). A rather homogeneous element distribution is expected for a diffusionless martensitic transformation, and these experiments indicate that this is indeed the case – despite the high temperatures reached during FLA (see next section).

Furthermore, it should be noted that a few film cracks are observed along $[100]_{MgO}$ (an example is given in Figure 9c). As shown by tilted-view SEM imaging of a FIB-prepared cross-section through such a crack, not only the Ni-Mn-Ga film and the Cr buffer are affected, but the crack goes very deep into the substrate (Figure 9d). All these observations indicate significant thermal stress. Detailed interpretations are given in Subsection 5.4.

4. **Simulation of temperature evolution**

To understand the influence of FLA on the martensitic microstructure, it is necessary to know the temperature evolution during and after FLA. For the interpretation of the previous experimental observations within the next section, three *mechanisms* to be considered will be derived in the following.

To obtain the temperature of the film and substrate, finite element calculations of the FLA process, which consider the optical and thermal properties of the Ni-Mn-Ga film and the MgO substrate were performed. (Details and material properties are provided in Subsection 7.2.) In *Figure 10*, the temperature evolution of the film surface is shown for the three key values of $E_D$ = 28, 55, and 83 J/cm$^2$. At $E_D$ = 55 J/cm$^2$, the surface temperature reaches a maximum of 597 K. This temperature is just below the transformation temperature of around 610 – 615 K, as determined by temperature-dependent optical microscopy. (Details are given in Supplementary Figure S7.) When considering that both experimental and simulated temperatures exhibit systematic errors, it can be concluded that with $E_D$ = 55 J/cm$^2$, the temperature is just at the threshold to induce a martensitic transformation within our films. The aspect that there is a threshold value of $E_D$ to induce an MT by FLA will be referred to as *mechanism 1*.

In addition, the calculated temperature evolutions give an idea of the cooling rate through the transition, which is determined by the slope of the cooling curve at the martensitic start temperature ($M_s$). A cooling rate of 20 K/ms is obtained for the highest energy density of $E_D$ = 83 J/cm$^2$. This cooling rate, which is dominated by heat conduction towards the cold backside, is orders of magnitude faster than that reachable by any conventional heating and cooling setup. The cooling rate, however, decreases with decreasing $E_D$, which is evident when considering the curve for $E_D$ =

55 J/cm$^2$. When assuming that this curve would just reach $M_s$, the slope at $M_s$ would be zero, which is equivalent to a zero cooling rate. For a martensitic transformation, the cooling rate is important since rapid cooling leaves less time for the formation of a martensitic microstructure. The reduced time with increasing energy density is considered as *mechanism 2*. However, this is only relevant if the energy density is sufficient to induce a martensitic transition (*mechanism 1*).

*Figure 10* also displays the calculated temperatures at the bottom of the MgO substrate ($T_{bottom}$), which are much lower compared to the film surface ($T_{top}$). This originates from the strong differences in their thickness, optical and thermal properties, as the thermal conductivity of the oxidic substrate is much lower than that of the metallic film and the substrate is much thicker than the film. The metallic film absorbs nearly all light, and due to the low thermal conductivity and the relatively large thickness of the MgO substrate, the heat transport towards the backside is rather limited, leading to a large temperature gradient. As a result, the film and the top of the substrate heat up much more than the bottom of the substrate. This temperature difference consists throughout the simulation time. The compressive stress can be quite large, since the hot metallic film also has a higher coefficient of thermal expansion than the colder oxidic substrate. The stress is not quantified here, since the elastic properties of martensitic films strongly depend on direction, phase, and temperature [27], and their pseudoelastic and plastic behaviour allows to compensate for a part of the stress. However, it is plausible to expect a higher thermal stress with increasing $E_D$. We consider thermal stress as *mechanism 3*, which ultimately depends on *mechanism 2*, since the time is too short to reach thermal equilibrium between film and substrate during FLA. In the following discussion, however, we can clearly differentiate between these two mechanisms.

## 5. Interpretations and discussions

In this section, the experimentally observed changes of the martensitic microstructure in Section 3 are interpreted based on the processes causing the formation of the hierarchical microstructure with five levels of twinning, as summarized in Section 2 and sketched in Figure 1a [19]. To this end, the three *mechanisms* derived from the simulations of the temperature evolution in Section 4 are considered: 1) MT occurs only above a certain threshold value of $E_D$. 2) With increasing $E_D$ above this threshold, less time is available for microstructure formation, and 3) with increasing $E_D$, an increasing compressive stress occurs in the film.

## 5.1 At the nanoscale (level 0...2): ordering and coarsening

The interpretation begins at level 0 at the nanoscale. The measurements of the lattice parameters of all three phases A, NM, and 14M (*Figure 4*a) reveal that they remain almost identical to those in the as-deposited state. Thus, FLA does not influence the MT itself and the fundamental building block of martensitic microstructure: the tetragonal NM unit cell. This observation is expected based on previous measurements [17], which demonstrated that an MT in this system can proceed even within several nanoseconds, i.e. much faster than during the FLA experiments. For the following interpretation, this is important, as the presence of the identical building blocks allows for the comparison with the known processes of microstructure formation.

At level 1, significant changes of the intensity fractions of A, NM, and 14M are observed, but only for $E_D \geqslant 55$ J/cm$^2$ (*Figure 4*b). Indeed, this threshold behaviour, as well as the threshold value, agrees very well with mechanism 1: Below the threshold $E_D$, no MT occurs, and accordingly, the phase fractions remain unchanged. A similar threshold behaviour is observed for several properties discussed in the following, and therefore, the background in these figures is shaded in grey up to the threshold value. Above the threshold $E_D$, an increase of NM and a decrease of 14M and A are observed. The transformation from 14M to NM can proceed by the coarsening of nano-twin boundaries, as sketched in *Figure 4*d [24]. After FLA, this process can be disturbed substantially. Since 14M is a $(2\bar{5})_2$-ordered arrangement of nano-TBs [21] and the process of ordering requires some time, the observed dominance of NM over 14M probably originates from the reduced time available during FLA (mechanism 2). However, stress also favours this transition, as known from bulk materials [32]. Therefore, mechanism 3 can also account for this observation. Furthermore, the observed disappearance of residual austenite indicates film stress, as stress favours an MT. Hence, at level 1, mechanisms 2 and 3 can both explain the experimental observations, but a differentiation, which one is more influential, cannot be made.

At level 2, the coherence length $D$ (*Figure 4*c) is considered. In epitaxial thin films, $D$ is primarily limited by the distance between TBs, which disrupt the lattice coherence. The different origins of $D$ from different Bragg reflections are sketched in *Figure 4*e. For NM martensite, an increase of $D$ (*Figure 4*c) is observed, as expected for coarsening of nano-TBs (as already observed for level 1 and described in the previous paragraph). For $D_{(400)NM}$, there is a moderate increase, directly reflecting the spacing of the nano-TBs. $D_{(004)NM}$ exhibits an even stronger increase. To understand this difference, it has to be considered that for tetragonal NM martensite, there are two different orientations for (004)$_{NM}$ with respect to TBs with (101)$_{NM}$ orientation. In addition to the $D_{(004)NM}$ illustrated in *Figure 4*e, an additional orientation of $D_{(004)NM}$ perpendicular to this sketch exists (which obviously cannot be sketched). This orientation is not limited by nano-TBs and accordingly, much larger coherence lengths are possible, in agreement with the experiments. Thus, this $D_{(004)NM}$ is limited by TBs at higher levels, and according to the following sections their spacing is indeed much larger than for nano-TBs. For 14M, the coherence length is limited by the distance of (a/b)$_{MM}$-TBs. $D_{(400)14M}$ and $D_{(040)14M}$ both rise slightly with $E_D$. Simultaneously, the phase fraction of 14M decreases. Probably only well-ordered 14M remains. To sum up, complementary to the increase of the NM phase fraction and the decrease of 14M at level 1, there is also a strong increase of the coherence lengths for all phases at level 2.

## 5.2 At the mesoscale (level 3): more and small nuclei

At level 3 at the mesoscale, the twinning period $\Lambda_{TB}$ originates from the width of a martensitic nucleus, as sketched in *Figure 1*a [20]. Up to a threshold $E_D$, no changes are observed, as expected by mechanism 1. This threshold of $E_D \leqslant 28$ J/cm$^2$ is a bit smaller than expected from the temperature calculations. This deviation should originate from the original experimental data since the former value is extracted from x-ray diffraction method which reaches the whole film, in contrast to this value extracted from SEM images, which only the sample surface are considered. As for the previous experiments, the same origin of this threshold behaviour is considered: a low $E_D$ is not sufficient to induce an MT, and accordingly, the martensitic microstructure remains in the as-deposited state. Above the threshold, a linear decrease in nucleus size with a further increase of $E_D$ is observed. As $\Lambda_{TB}$ is controlled by the autonucleation, which creates further nuclei from the primary nucleus, we can attribute this observation to mechanism 2. A fast undercooling at higher $E_D$ increases the driving energy for autonucleation and accordingly more and smaller nuclei form, which results in a reduced $\Lambda_{TB}$. Thus, mechanism 2 – less time available for processes of microstructure formation – explains the observed changes of $\Lambda_{TB}$ in dependence of $E_D$. The mesoscale is decisive for most applications, e. g. mesoscopic twin boundaries are easily mobile [33], which is essential for magnetic actuation. Our FLA experiments identify that cooling within the millisecond range is suitable to control the nucleation behaviour and, thus, the mesoscopic twin boundary period.

### 5.3 At the macroscale (level 5): nucleation and stress-induced reorientation

At level 5, two types of macroscopic TBs were examined (Subsection 3.3). First, for the type-X colony size, a threshold behaviour was observed (*Figure 6*b), as expected by mechanism 1. For $E_D \geqslant 42$ J/cm$^2$, there is a strong increase of the total colony number by nearly three times, which, in particular, originates from the increase of colonies smaller than 10 µm$^2$. As each colony originates from at least one primary nucleus, which creates a colony by spontaneous nucleation [19], this observation is expected by classical nucleation theory: fast cooling allows for stronger undercooling before nucleation starts, which increases the driving force foe nucleation. Accordingly, smaller and more primary nuclei form. This results in smaller colonies, which agrees with our observation.

Second, the fractions of type-X and -Y were examined. With two independent methods, SEM imaging (*Figure 7* and *Figure 8*) and a quantitative analysis of an XRD pole figure (Figure S5), a linear increase of type-X with increasing $E_D$ is revealed. In particular, no threshold is observed; rather, redistribution between both types already starts at the lowest $E_D$ applied, which rules out mechanisms 1 and 2. Mechanism 3 is therefore considered the cause, as the thermal stress increases with $E_D$ without any threshold. Indeed, type-X and -Y are expected to respond differently to stress [20]. These differences originate from the Schmid factor, which describes how stress acts on a twinning plane [34]. When the relative orientation of twinning plane and stress is perpendicular or parallel, this factor is zero, meaning that stress has no influence on twinning. The Schmid factor is maximum at a relative orientation of 45°, indicating that this orientation of twin boundaries can easily accommodate stress Indeed, type-X and type-Y differ in their TB orientation with respect to the film plane, where biaxial thermal stress is acting. The type-Y TB is aligned perpendicular to the substrate plane, meaning stress is expected to have no influence. In contrast, the type-X TB is inclined by 45°, making it more favourable under stress. In conclusion, the reorientation from type-Y to type-X with increasing $E_D$ is attributed to mechanism 3.

### 5.4 At the macroscale after large over-heating: deformation and cracking

Thermal stress is a known mechanism during FLA [18]. Our observation of wrinkles (hillocks and valleys) and cracks at maximum energy density (Subsection 3.4) suggests the following scenario, which is explained by mechanism 3: during FLA, most of the heat is absorbed by the metallic film, causing a significant temperature difference between the top and bottom of the substrate. This results in compressive stress at the top of the substrate. Due to the pseudoplastic properties of the shape memory alloys, the film remains intact, while the brittle MgO substrate is already deformed plastically. Upon cooling to ambient temperature, both the film and substrate are expected to return to their original states. However, the plastic deformation of the substrate, caused by the slightly increased film thickness in the hillocks with bumps at the interface to the substrate necessitates the formation of valleys with reduced film thickness to compensate. This explains the simultaneous occurrence of both hillocks and valleys.

With a further increase of $E_D$ to 83 J/cm$^2$, cracks occur within the substrate surface (Figure 9c-d). The substrate breaks with the expected clear brittle cuts, while the film exhibits plastic deformation, as indicated by its increased length and the typical fractured surface with an angle of 45° (*Figure 7*d). The orientation of all these features is along ⟨100⟩ directions of the MgO, which is expected since MgO single crystals easily cleave along {100} planes, and the thin film follows these cleavage path. It's important to note that these effects also exhibit an energy density threshold. However, this threshold is not related to mechanism 1, but to the ultimate thermal stress of MgO, which is a completely different type of threshold.

### 6. Conclusion and outlook

In this work, we studied the formation of the martensitic microstructure in epitaxial Ni-Mn-Ga films under a fast transformation induced by flash lamp annealing (FLA). The use of epitaxial films as a model system and the combination of various complementary methods allows to quantify microstructural changes at almost all levels of twinning. Three mechanisms which control the microstructure formation in dependence on the energy density $E_D$ were identified: 1) martensitic transformation (MT) occurs only above a certain threshold $E_D$. 2) With increasing $E_D$, less time is available for microstructure formation. 3) With increasing $E_D$, an increasing compressive stress occurs in the film. As summarized in *Figure 1*b, these three mechanisms explain the observed changes in microstructure, which impact almost all levels of twinning. Our analysis provides an understanding of the dynamics of the different processes governing the formation of the hierarchically twinned microstructure. Specifically, the reduced time available during FLA (mechanism 2) hinders the ordering of nano-twin boundaries (level 1 and 2), which results in smaller nuclei (level 3), and reduces the colony size (level 5). Additionally, thermal stress during FLA (mechanism 3) induces a reorientation from type-X to -Y (level 5) and causes wrinkling and cracking at high energy densities.

Given that most functional properties of shape memory alloys depend on the martensitic microstructure, our approach marks a significant step toward disentangling the various processes and their dynamics. This fundamental understanding is crucial for future devices, which are expected to be faster and smaller to achieve better performance. This quantification can be expanded to FLA experiments at different pulse durations to conduct a time-temperature study of the dynamics during the formation of a martensitic microstructure. Such studies will provide benchmarks for simulations of martensitic transformation across different time and length scales.

## 7. Methods

### 7.1 Film deposition and transformation temperature

As described in our previous works [28, 29], Ni-Mn-Ga films were grown epitaxially on single-crystalline MgO (100) substrates (Crystec GmbH, Germany) by a DC magnetron sputter deposition tool from Bestec GmbH, Germany. Films with 500 nm thickness were deposited at 400 °C within the austenitic state. To adjust the lattice mismatch and enhance the cohesion between film and substrate, a 20-nm-thick Cr buffer layer was deposited at 300 °C below the Ni-Mn-Ga film [29]. The nominal target stoichiometry is $Ni_{48}Mn_{32}Ga_{20}$. The substrate holder rotated continuously during deposition to ensure a uniform thickness and composition. After deposition, the films cooled down to room temperature and transformed into the martensitic state. The transformation temperature was determined using an optical microscope (Zeiss GmbH, Germany) with a heater. (Details are shown in Supplementary Figure S7.)

### 7.2 Flash Lamp Annealing (FLA)

To induce a fast MT in the as-deposited film, an FLA tool from Rovak GmbH, Germany was employed. FLA is a non-equilibrium annealing method to achieve a fast thermal treatment of solid surfaces within hundreds of microseconds up to tens of milliseconds [15]. A basic scheme is sketched in Supplementary Figure S8. The FLA chamber consists of one or several xenon lamps at the top, which heat the sample by radiation.

The reflector is designed to ensure a homogeneous temperature of the whole sample. The key parameter in FLA is the energy density per pulse ($E_D$), given in J/cm$^2$, which is an integral measure over the optical spectrum and the pulse duration. In the present work, one as-deposited thin film sample was cut into four identical pieces, each having a size of 0.5 × 0.5 cm$^2$. Seven of these pieces were heated individually with different $E_D$ values ranging from 14 to 83 J/cm$^2$ in a continuous N$_2$ flow of about 0.1 bar within the chamber. A single pulse of 3 ms duration was applied.

To determine the temperature evolution within the film and the substrate during and after FLA, simulations were performed by solving the one-dimensional heat equation using COMSOL Multiphysics. The energy density $E_D$, pulse shape, and flash lamp spectrum were obtained from calibration measurements [30]. The optical constants of Ni$_2$MnGa [31] and MgO [32], and the thermodynamic properties of Ni$_2$MnGa [33] and MgO [34] were taken from the literature. The involved values required for simulation are listed in Supplementary in Table S1. Three exemplary temperature evolutions are shown in *Figure 10*. During the 3-ms flash pulse, the temperature of the film surface ($T_{top}$) increases sharply from room temperature to a maximum temperature. Simultaneously, the bottom-side temperature of the substrate ($T_{bottom}$) remains close to ambient temperature. If $T_{top}$ exceeds the transformation temperature, the film transforms to austenite. After the pulse, heat transfers from the film to the substrate due to the large difference between $T_{top}$ and $T_{bottom}$. The substrate with its relatively large heat capacity serves as a "heat sink", and the film cools down and transforms back to martensite. The heating and cooling rates are estimated to be in the magnitude of 10…100 K/ms. The simulated temperature profiles may have some uncertainty, because bulk properties of Ni-Mn-Ga were taken from the literature and used for the thin film.

### 7.3 Characterization

To characterize the crystal structures of the martensitic types and variants, reciprocal space maps (RSM) within a χ range from -5° to 5° were recorded in 2D mode on a SmartLab diffractometer (Rigaku, Japan) at room temperature. The instrument uses a parallel beam of Cu-Kα radiation and is equipped with a HyPix-3000 two-dimensional semiconductor detector. The purpose of covering χ from -5° to 5° is to capture martensitic variants that are slightly tilted. In Supplementary Figure S1, a side-view sketch of the diffraction geometry is given. Furthermore, it is described how the 2D

RSM measurements are performed and important phase data are extracted. To examine the influence of $E_D$ on the martensitic microstructure, a Sigma 300 scanning electron microscope (SEM) (Zeiss GmbH, Germany) with a secondary-electron (SE) detector, operated at an accelerating voltage of 5 kV, was used. Properties of twinning period, colony and variant fractions were extracted from the recorded SEM images by processing with GNU Image Manipulation Program (GIMP) and MATLAB (scripts are available at https://doi.org/10.14278/rodare.3135).

To investigate the wrinkles and cracks on the film surface in more detail, cross-sections were prepared using a Helios 5 CX focused ion beam (FIB) / SEM device (Thermo Fisher, Eindhoven, Netherlands). For better contrast and to protect the film surface, first, a carbon cap layer was deposited with electron-beam-assisted precursor decomposition. (In Figure 9b, the corresponding layer is denoted with "eC".) Second, a Pt double layer was deposited beginning with electron-beam-assisted and subsequently followed by Ga-FIB-assisted precursor decomposition. (In Figure 9b, the corresponding single layers are denoted with "ePt" and "iPt", respectively.) The cross-section itself was prepared using a 30-keV Ga-FIB with adapted currents and imaged at a tilt angle of 52° with the SEM of the Helios 5 CX operated at an accelerating voltage of 2…5 kV. Cross-sectional TEM lamella preparation was done by in situ lift-out also using the Helios 5 CX FIB/SEM device. To protect the film surface, a carbon cap layer was deposited beginning with electron-beam-assisted and subsequently followed by Ga-FIB-assisted precursor decomposition. Afterwards, the TEM lamella was prepared using a 30-keV Ga-FIB with adapted currents. Its transfer to a 3-post copper Omniprobe lift-out grid was done with an EasyLift EX nanomanipulator (Thermo Fisher). To minimize sidewall damage, Ga ions with only 5-keV energy were used for final thinning of the TEM lamella to electron transparency. Bright-field TEM imaging was performed with an image-$C_s$-corrected Titan 80-300 microscope (FEI, Eindhoven, Netherlands) operated at an accelerating voltage of 300 kV. High-angle annular dark-field scanning transmission electron microscopy (HAADF-STEM) imaging and spectrum imaging analysis based on energy-dispersive X-ray spectroscopy (EDX) were performed at 200 kV with a Talos F200X microscope equipped with an X-FEG electron source and a Super-X EDX detector system (FEI). Prior to (S)TEM analysis, the specimen mounted in a double-tilt low-background holder was placed for 8 s into a Model 1020 Plasma Cleaner (Fischione, Export, PA, USA) to remove potential contamination.

## Supporting Information

Supporting Information is available at XXXXXX.

## Author contributions

Y.G., L.R., and S.F. conceived the experiments. With the support of F.G. and K.L., Y.G. conducted most of the experiments, wrote the MATLAB scripts, analysed the data and wrote the first version of the manuscript. S.K. deposited the films and determined the transformation temperatures. R.H. performed the (S)TEM analysis. L.R. and S.Z. supported the FLA and VSM measurements. L.R. conducted the FLA process and simulated the FLA temperature evolution. S.F. conducted the data interpretation. All authors contributed with descriptions of their results, manuscript corrections, and agreed to the published version of the manuscript.

## Authors' ORCIDs

Yuru Ge 0000-0002-3977-2505

Fabian Ganss 0009-0003-0366-9690

Klara Lünser 0000-0003-3309-7948

Satyakam Kar 0000-0003-1352-8026

René Hübner 0000-0002-5200-6928

Shengqiang Zhou 0000-0002-4885-799X

Lars Rebohle 0000-0002-8066-6392

Sebastian Fähler 0000-0001-9450-4952

## Acknowledgements

The authors thank Thomas Naumann and Olav Hellwig (HZDR) for support with the SmartLab, Andreas Worbs (HZDR) for FIB support, Lothar Bischoff and Nico Klingner (HZDR) for SEM support, Thomas Schumann (HZDR) for FLA support, and Rudolf Schäfer (Leibniz IFW Dresden) for support with the optical microscope. Y.G. thanks Jürgen Lindner (HZDR), Andreas Undisz (TU Chemnitz), Jürgen Faßbender (HZDR), and Gianaurelio Cuniberti (TU Dresden) for their support and helpful discussions. Furthermore, the use of the HZDR Ion Beam Center TEM facilities and the funding of TEM Talos by the German Federal Ministry of Education and Research (BMBF, grant No. 03SF0451) in the framework of HEMCP are acknowledged.

**Conflict of Interest**

The authors declare that they have no known competing financial interests or personal relationships that could have appeared to influence the work reported in this paper.

**Data Availability Statement**

The data that support the findings of this study are openly available in RODARE at https://doi.org/10.14278/rodare.3135.

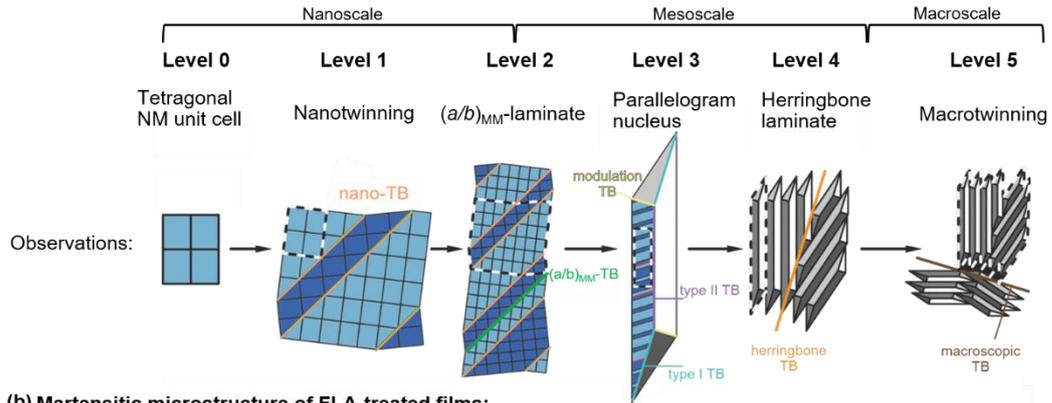

Figure 1: (a) Overview of the hierarchical martensitic microstructure after slow cooling as reference state (adapted from [16]). The microstructure consists of nested building blocks and twin boundaries (TBs). There are five microstructure levels, starting from the tetragonal non-modulated martensite (NM) cell as initial building block. In each subsequent level, the next building block is obtained by combining building blocks from the previous level by twinning (TBs shown in different colours). These five levels cover all length scales, from the nano- up to the macroscale. (b) Influence of FLA on this hierarchical twinning. This table summarizes the observations of chapter 3, which are interpreted in chapter 5.

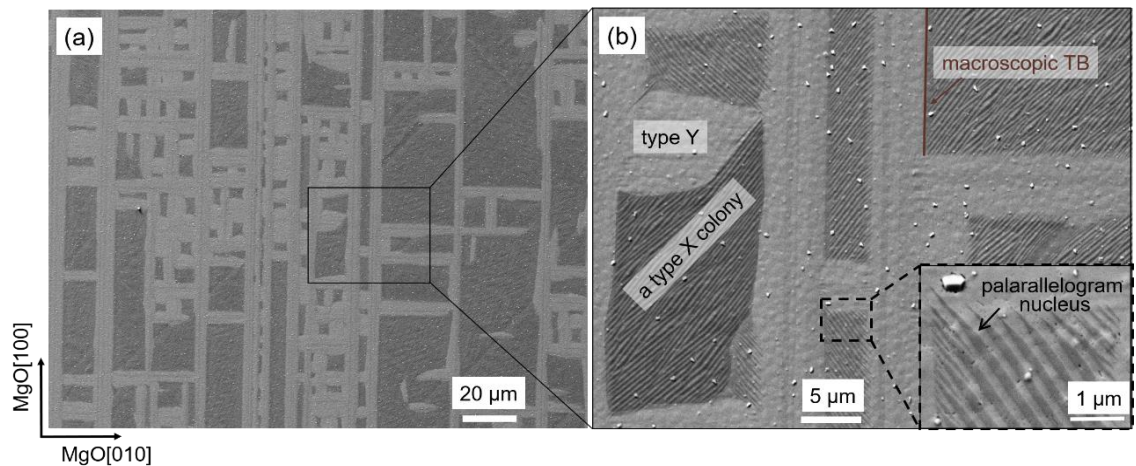

Figure 2: Hierarchical martensitic microstructure in the as-deposited state. (a) An SEM overview showing two distinct features of type-X (dark area) and type-Y (bright area), which are connected by macroscopic TBs. (b) A zoomed-in image to illustrate where type-X, type-Y, and macroscopic TBs are present. A further magnification in the bottom right corner (dashed lines) reveals the parallelogram-shaped nuclei. (Note: to better illustrate the microstructure, these images were captured from a 2000 nm thick as-deposited film, used here only as examples)

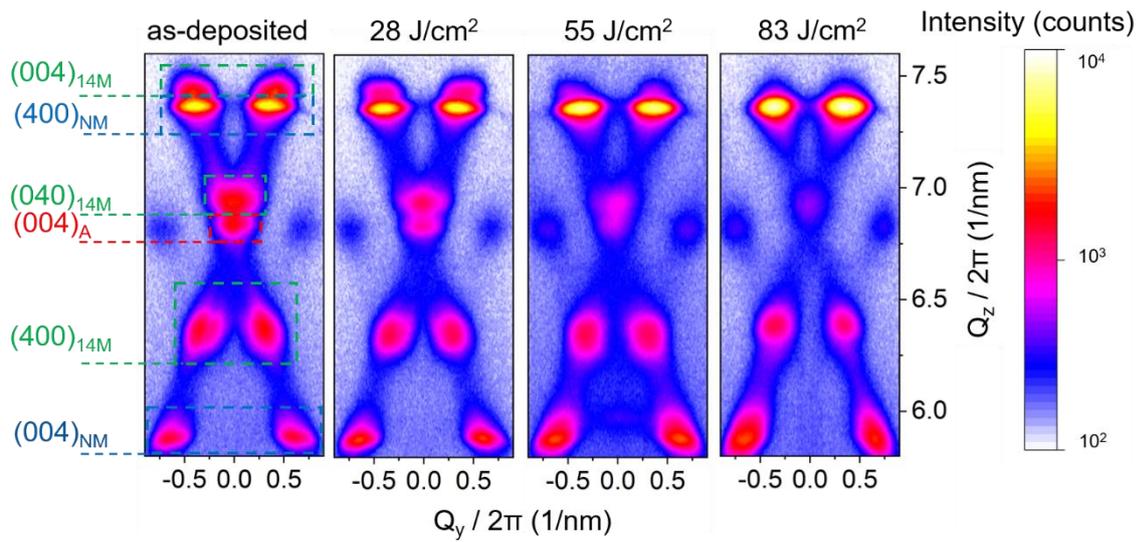

Figure 3: RSMs of the as-deposited film and the FLA-treated films with $E_D$ = 28, 55, and 83 J/cm² show that the residual austenite is reduced by FLA and the transition of 14M to NM occurs with increasing $E_D$.

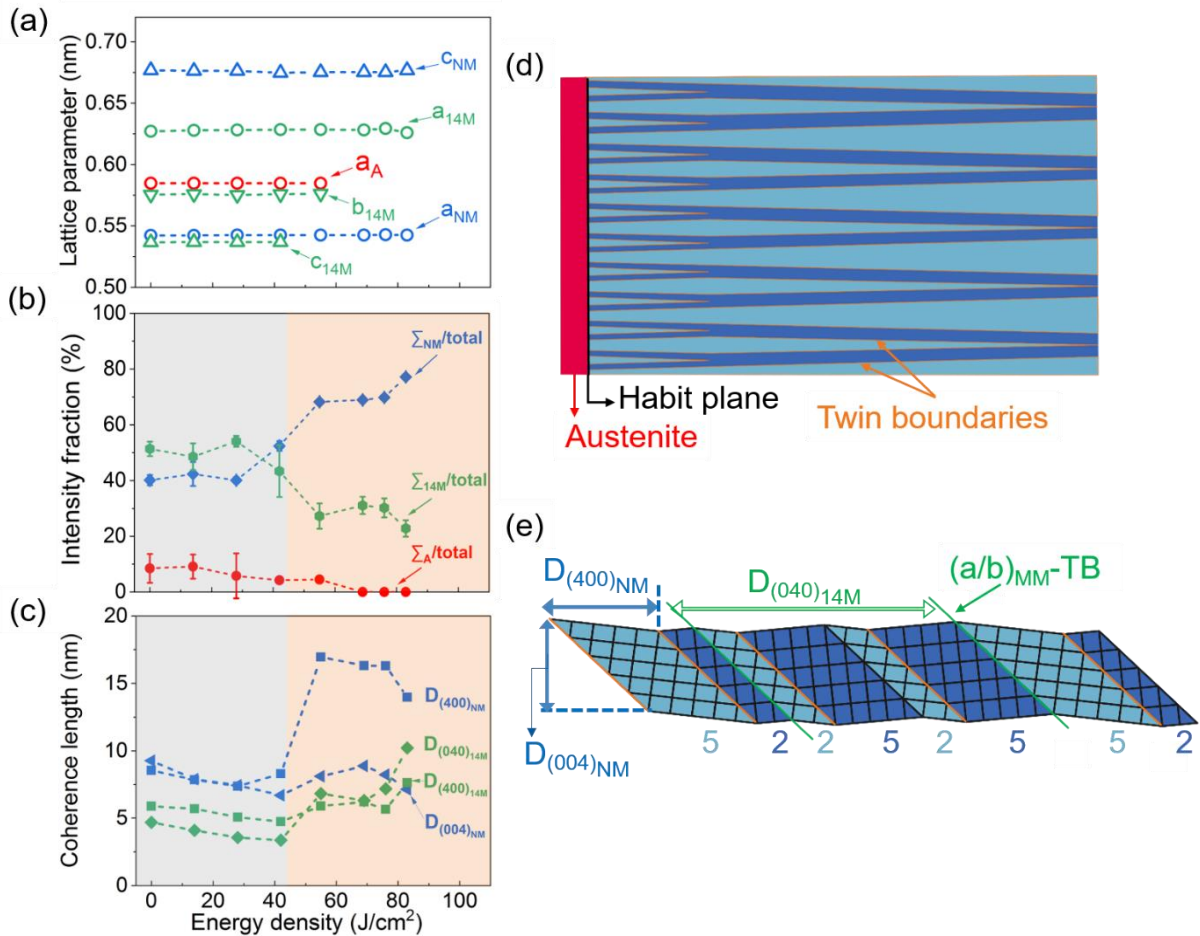

Figure 4: To understand the impact of $E_D$ on the atomic structure and microstructure, three characteristic phase properties were extracted from the RSMs: (a) lattice parameters, (b) intensity fractions, and (c) coherence lengths $D$. (d) By FLA, a transition from 14M to NM occurs, which proceeds by coarsening of nano-TBs, sketched here (adapted from [21]). (e) Sketch of the different coherence lengths D within a twinned martensitic microstructure (level 2 from Fig. 2a). Note that there are two $D_{(004)NM}$ – the one perpendicular to this sketch cannot be shown and is not affected by nano-TBs. Based on this sketch, the influence of FLA on the different $D$ values is discussed in section 5.1.

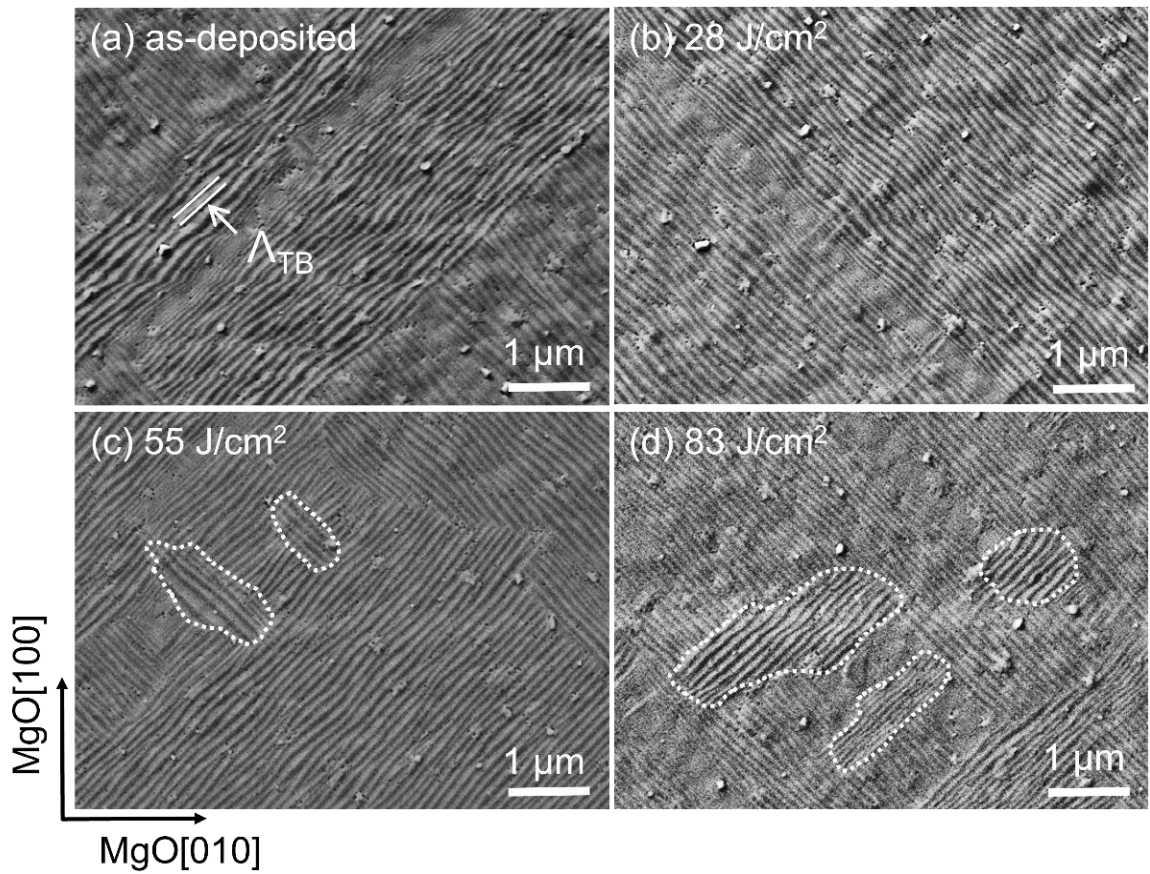

Figure 5: SEM micrographs of (a) the as-deposited film and the FLA-treated films with $E_D$ = (b) 28, (c) 55, and (d) 83 J/cm$^2$ show a decreasing twinning periodicity $\Lambda_{TB}$ and the occurrence of small type-X colonies, marked with dashed contours. The figure edges are parallel to [100]$_{MgO}$ and [010]$_{MgO}$ for all figures.

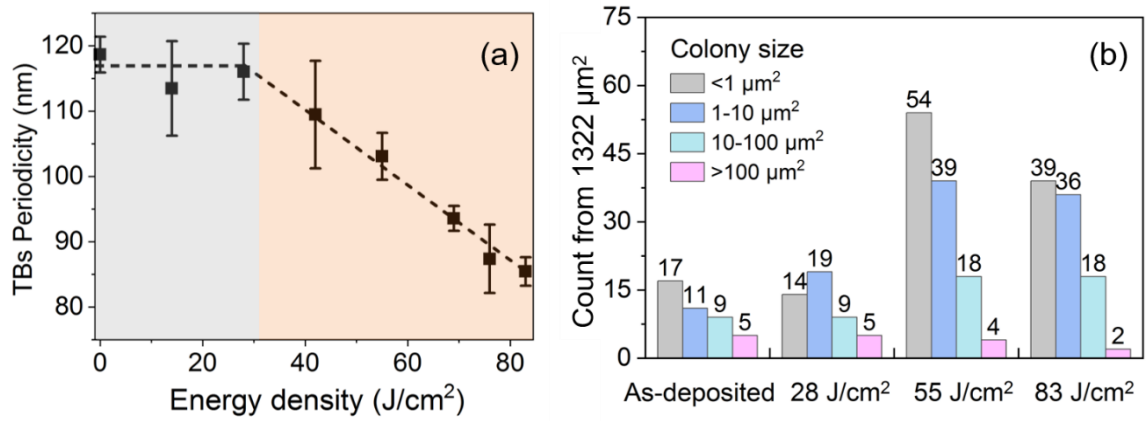

Figure 6: (a) Type-X TB periodicity ($\Lambda_{TB}$) as obtained from SEM images as the function of $E_D$. $\Lambda_{TB}$ remains the same for $E_D \leqslant 28$ J/cm² and decreases linearly for higher values. (b) Statistics on colony sizes obtained from several SEM micrographs with a total area of 1322 µm² shows that smaller colonies occur for $E_D > 28$ J/cm².

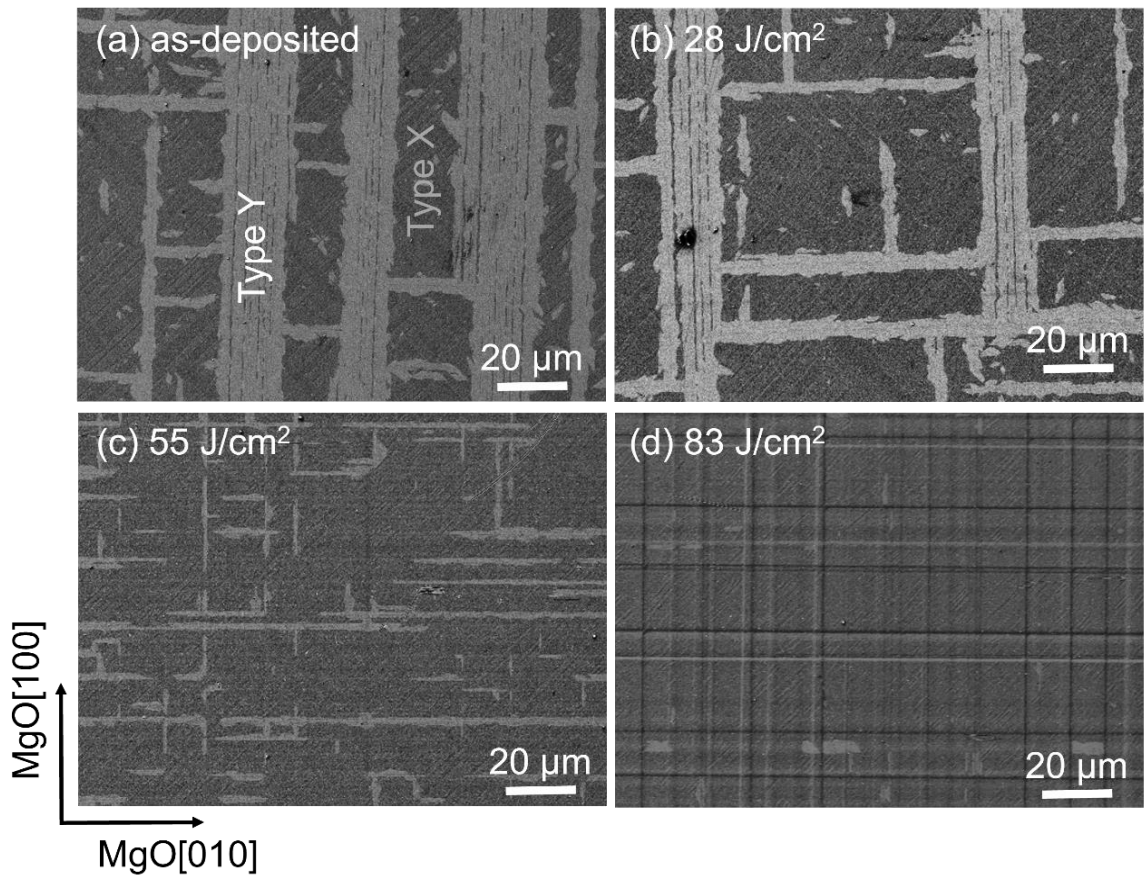

Figure 7: (a-d): SEM micrographs at low magnification of (a) the as-deposited film and the FLA-treated films with $E_D$ = (b) 28, (c) 55, and (d) 83 J/cm² show an increasing fraction of type-X variants (dark region) and correspondingly a decrease of type-Y variants (bright region) with increasing $E_D$. At $E_D$ = 83 J/cm², the film exhibits wrinkles along the substrate orientations [100]$_{MgO}$ and [010]$_{MgO}$. The figure edges are parallel to [100]$_{MgO}$ and [010]$_{MgO}$ for all figures.

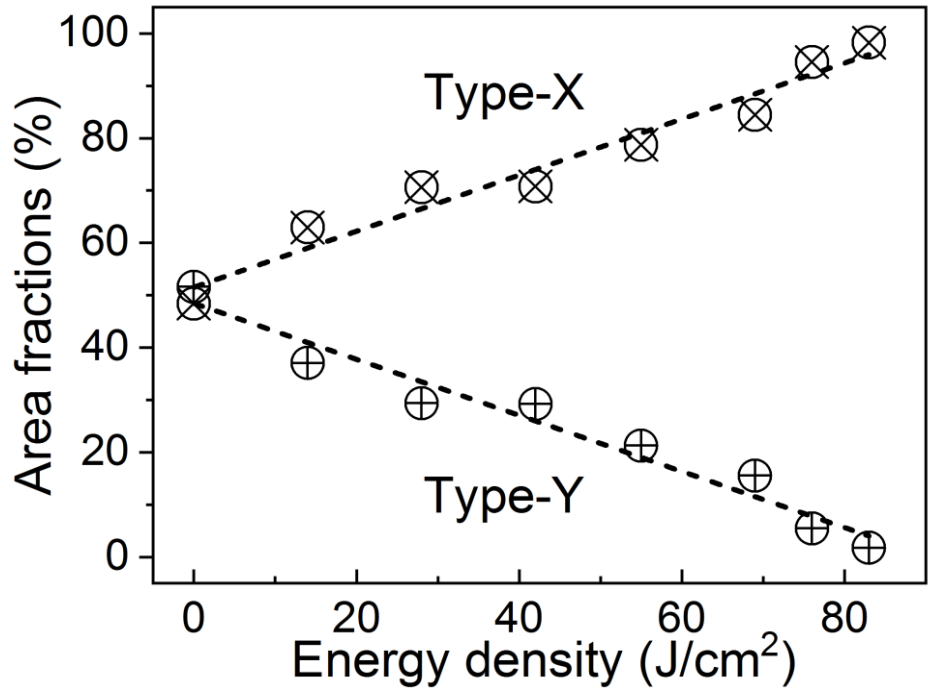

Figure 8: (a): Pixel fractions of type-X and type-Y were extracted from counting pixels of 5 SEM images at low magnification based on contrast thresholds. With increasing $E_D$, the samples are dominated by type-X.

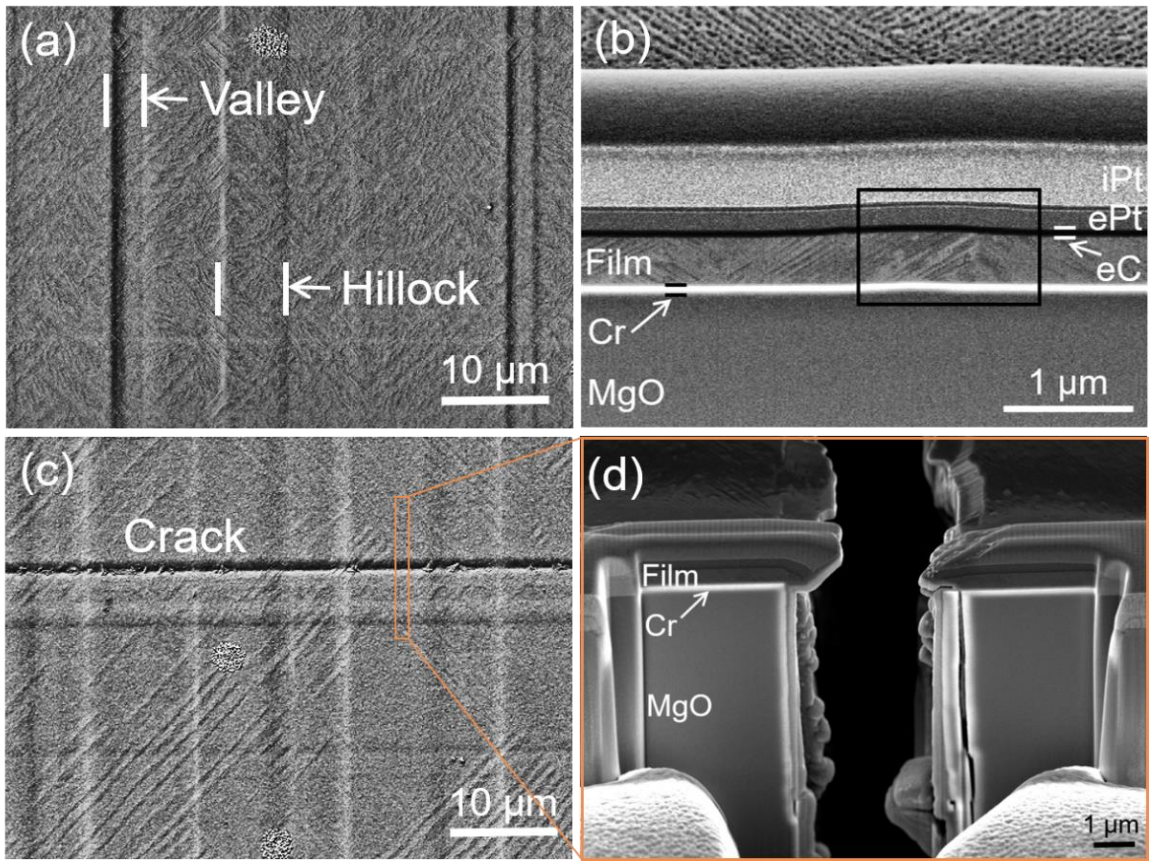

Figure 9: Detailed analysis of the dark and bright traces (in Figure 7d) parallel to <100>$_{MgO}$ occurring at $E_D$ = 83 J/cm$^2$: (a) SEM imaging reveals two features on the film surface: valleys and hillocks. (b) A FIB-prepared cross-section cut through one narrow hillock examined by tilted-view SEM imaging shows a slight increase in film thickness with a bump at the interface, but no significant delamination is observed. (c) In a few regions, additional cracks are observed on the surface by SEM imaging. (d) A FIB-prepared cross-section through one crack shows that the film, the Cr buffer, and even the MgO substrate are all cracked. Both observations are indications for a large thermal stress, as discussed within the main text.

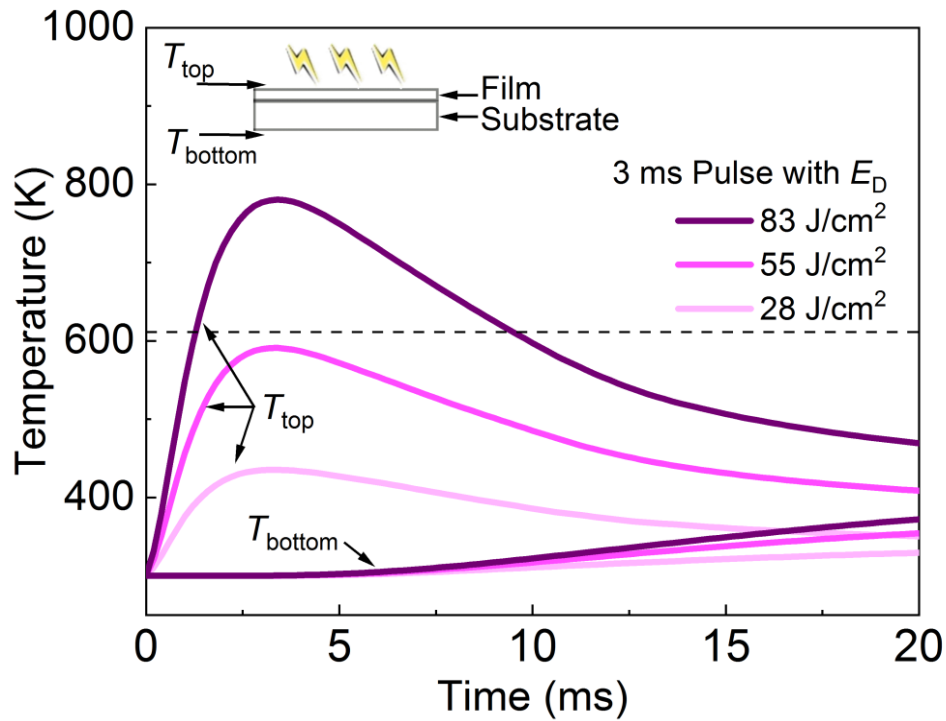

Figure 10: Simulated temperature evolutions of a 500-nm-thick Ni-Mn-Ga film grown on a 1-mm-thick MgO substrate for a 3 ms flash pulse with $E_D$ = 28, 55, and 83 J/cm² applied at an ambient temperature of 300 K. The upper and lower curves with the same colour are the temperature profiles of the top and the bottom side of the sample, respectively. The martensitic start temperature $M_s$ = 610 K of the film is marked by a horizontal dashed line. The temperature rates on heating and cooling are faster if a larger $E_D$ is applied.